\documentclass[12pt]{iopart}

\usepackage{iopams}
\usepackage{setstack}

\usepackage[dvips]{graphicx}

\begin{document}

\title[$\Lambda$(1520) production in d+Au collisions at RHIC]{$\Lambda$(1520) production in d+Au collisions at RHIC}

\author{Christina Markert for the STAR collaboration}

\address{\dag\ Physics Department, Kent State University, Kent, OH 44242,
USA}

\begin{abstract}
Recent results of $\Lambda$(1520) resonance production in d+Au
collisions at $\sqrt{s_{\rm NN}} = $ 200 GeV are presented and
discussed in terms of the evolution and freeze-out conditions of a
hot and dense fireball medium. Yields and spectra are compared to
results from p+p and Au+Au collisions. The
$\Lambda$(1520)/$\Lambda$ ratio in d+Au collisions ratio is
consistent with the ratio in p+p collisions. This suggests a short
time for elastic interactions between chemical and thermal
freeze-out. One can conclude that the interaction volume in d+Au
collisions is small.
\end{abstract}

\section{Introduction}

In heavy ion collisions an extended hot and dense medium is
created. The properties (mass, width, momentum distribution,
yield) of the produced resonances depend on the fireball
conditions of temperature, pressure and lifetime. During the
fireball expansion the short lived resonances and their hadronic
decay daughters may interact with the medium. Two freeze-out
surfaces can be defined, chemical and thermal, representing the
conditions when inelastic and elastic interactions cease
respectively. In a dynamically evolving system produced resonances
decay and may by regenerated. Hadronic decay daughters of
resonances which decay inside the medium may also scatter with
other particles from the medium. This results in a signal loss,
because the reconstructed invariant mass of the decay daughters no
longer matches that of the parent. The strength of the
rescattering and regeneration via the inverse decay processes for
resonances and their decay particles which are dominant after
chemical but before the kinetic freeze-out depend on individual
cross sections. These interactions can lead to changes of the
reconstructed resonance yields, momentum spectra, widths and
masses \cite{ble02}. Rescattering will decrease the measured
resonance yields while regeneration will increase them. For small
system sizes such as the one produced in p+p and d+Au collisions
we would expect a short or even no interaction time of hadrons
since the interaction volume is very small. In order to try to
understand the medium effect in a d+Au system during the evolution
and expansion of the fireball, we compare resonance yields and
spectra to elementary p+p collisions, and system size and time
extended Au+Au collisions.

\section{Resonance Reconstruction}

\begin{figure}[htb]
 \centering
 \vspace{0.5cm}
 \includegraphics[width=0.8\textwidth]{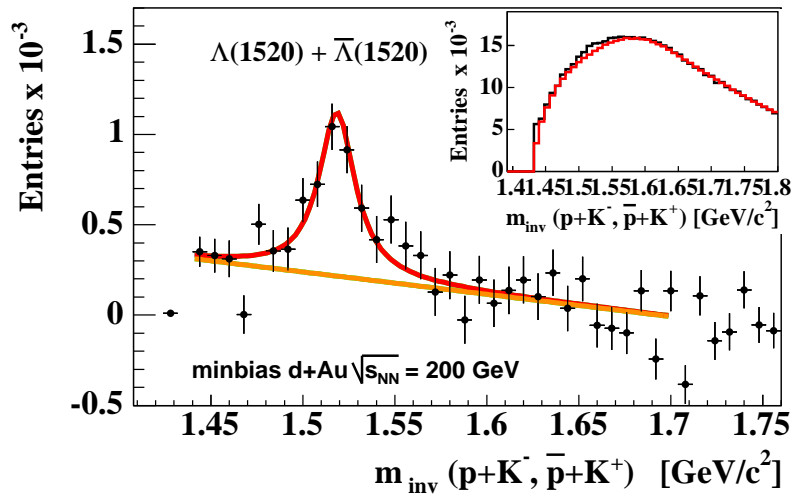}
\caption{Invariant mass distributions in d+Au collisions at
$\sqrt{s_{\rm NN}}=200\;\rm GeV$ before (inset) and after
mixed-event background subtraction for
 $\rm \Lambda^{*}\rightarrow p+K^{-}$.}

 \label{invmass}
\end{figure}

 The $\Lambda(1520)$ resonance is determined by the invariant mass
 reconstruction using the p and K$^{-}$ decay daughters. The decay
candidates are identified by their energy loss (dE/dx) in the STAR
{\it Time Projection Camber} (TPC). The resonance signal is
obtained by the invariant mass reconstruction of each daughter
combination and the subsequent subtraction of the combinatorial
background calculated via a mixed event technique.
Fig~\ref{invmass} shows the invariant mass spectrum before
(insert) and after background subtraction. The trigger selection
is set to a minimum bias trigger.

\section{Resonance Spectra and Yields}

\begin{figure}[t!]
\begin{minipage}[b]{0.5\linewidth}
 \centering
\includegraphics[width=1\textwidth]{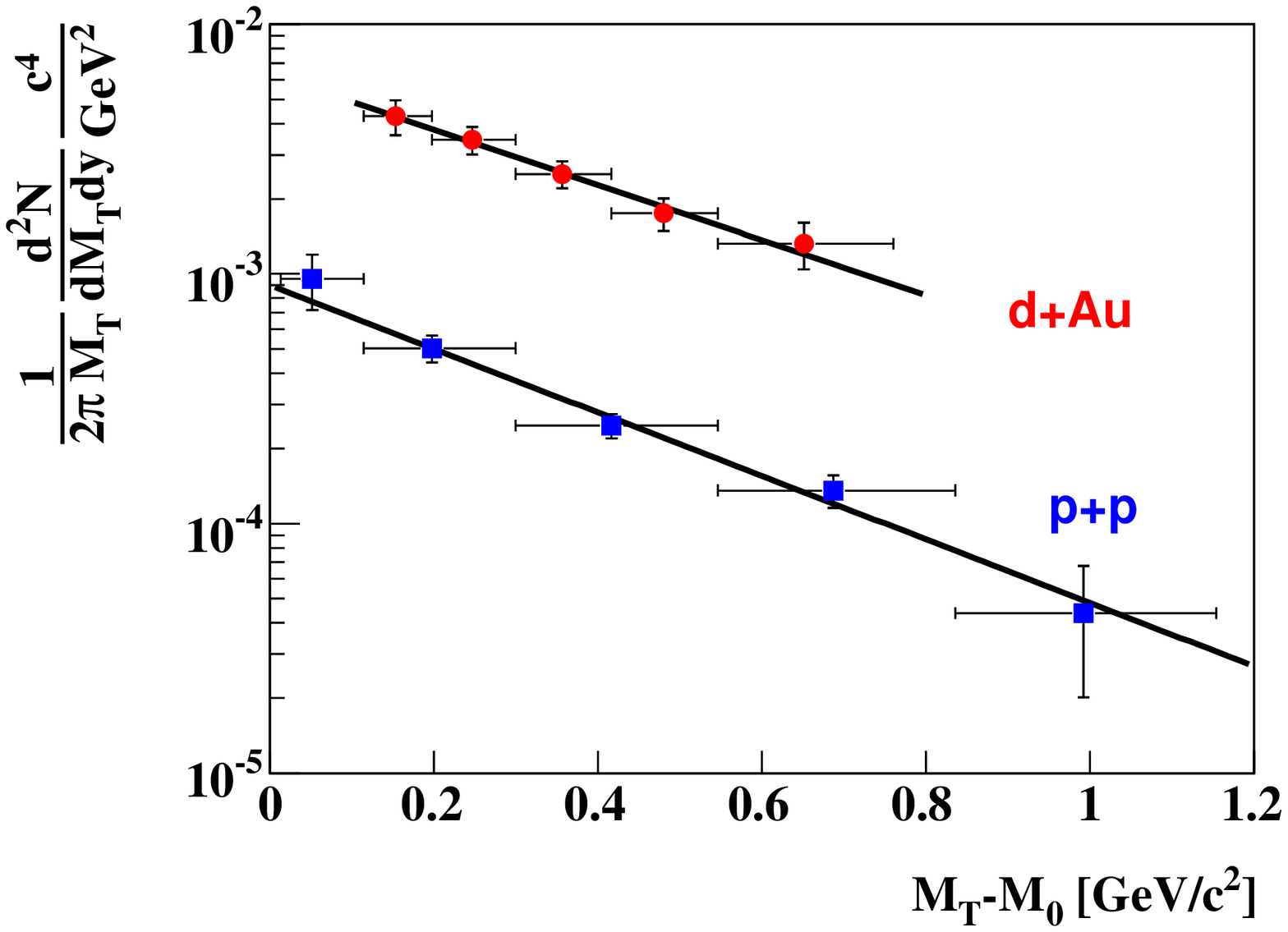}
\vspace{-0.9cm}
 \end{minipage}
 \begin{minipage}[b]{0.5\linewidth}
 \centering
\includegraphics[width=1\textwidth]{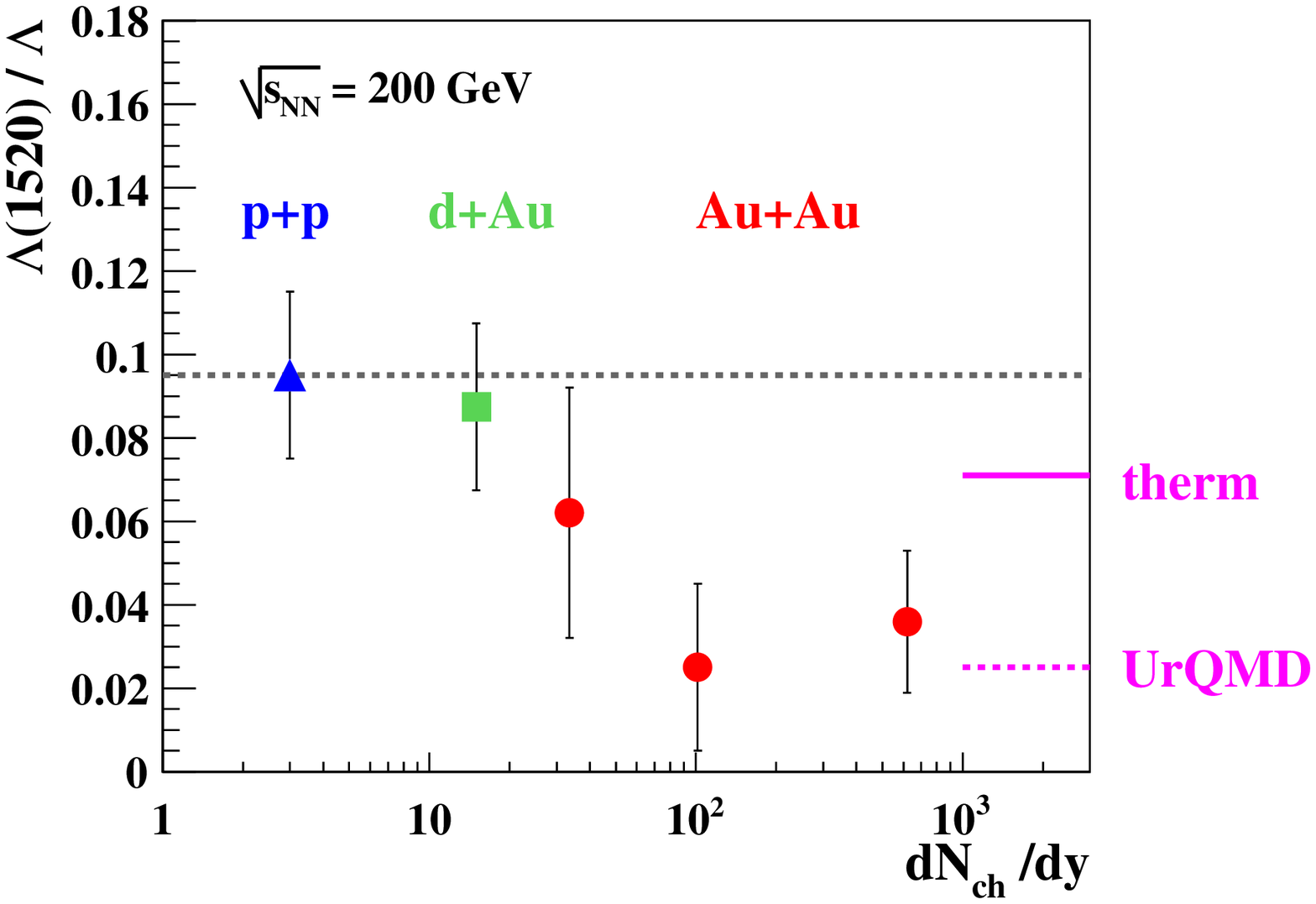}
\end{minipage}
 \caption{Left: The transverse mass spectra for $\Lambda$(1520)
  in p+p, d+Au collisions at $\sqrt{s_{\rm NN}}=
200 \;\rm GeV$. Right: $\Lambda$(1520)/$\Lambda$ ratio for p+p,
d+Au and Au+Au collisions at $\sqrt{s_{\rm NN}} = $ 200 GeV at
mid-rapidity. Statistical and systematic errors are included.
Thermal model and microscopic model predictions (UrQMD) for
$\Lambda$(1520)/$\Lambda$ in central Au+Au systems are shown.
Statistical and systematic errors are included.}
 \label{pt}
\end{figure}

The $\Lambda$(1520) transverse mass spectra at mid-rapidity are
shown in Fig~\ref{pt} (left) for minimum bias p+p and d+Au
collisions. The yield is obtained by the extrapolation and
integration of the momentum distribution. For Au+Au collisions an
inverse slope parameter has to be assumed (T~=~350~MeV) due to the
small signal. Yield and $\langle$p$_{\rm T}$$\rangle$ comparisons
are shown in Table~\ref{table}. In order to compare different
collision systems we normalize the yield to the yield of the
corresponding measured ground state particle. Fig.\ref{pt} (right)
shows the $\Lambda$(1520)/$\Lambda$ ratios versus number of
participants for different collision systems. This ratio remains
constant in p+p and d+Au collisions and decreases in Au+Au
collisions with increasing centrality \cite{mar04qm,mar03,gau04}.
The comparison to microscopic (UrQMD \cite{ble02}) and thermal
(\cite{pbm01,flo04}) model predictions for Au+Au collisions
indicates a hadronic interaction phase were re-scattering and
regeneration may occur, which can not be described by the thermal
model. Elastic interactions after chemical freeze-out happen only
in an extended reaction volume. In d+Au collisions we would not
expect an extended reaction volume and therefore the
$\Lambda$(1520)/$\Lambda$ ratio is expected to be similar to p+p
collisions, which is in agreement with the data. Based on the data
a lower limit of $\Delta\tau$ $>$ 4~fm/c can be placed on the
interaction time in central Au+Au collisions
\cite{tor01,mar02,tor01a}. The interaction time here is defined as
the time between chemical and kinetic freeze-out. As the
$\Lambda$(1520)/$\Lambda$ in d+Au is in agreement with the thermal
model and the value for p+p, where one expects a very short time
($\Delta\tau$ $\approx$~0~fm/c) for a hadronic phase, we can also
conclude a $\Delta\tau$ $\approx$~0~fm/c for the elastic
interaction hadronic phase in a d+Au collision.


\begin{table}
\centering \vspace{+0.5cm}
\begin{tabular}{|l|c|c|c|} \hline
  & T [MeV] & $\langle$p$_{\rm T}$$\rangle$ [MeV/c] & $(dn/dy)|_{y=0}$ \\ \hline \hline
\hline ($\bar{\Lambda}^{*}$+$\Lambda^{*})/2$ $_{\rm p+p}$ & 338 $\pm$ 42  & 1.08 $\pm$ 0.09 & 0.0034 $\pm$ 0.0006 \\
\hline ($\bar{\Lambda}^{*}$+$\Lambda^{*})/2$ $_{\rm d+Au}$ & 382 $\pm$ 75  & 1.17 $\pm$ 0.15 & 0.0149 $\pm$ 0.0014 \\
\hline $\Lambda ^{*}$ $_{\rm Au+Au}$ & fixed T=350 & -------- & 0.58 $\pm$ 0.28 \\
\hline
\end{tabular}
\caption{Temperature , yield, and $\langle p_{T}\rangle $ from an
exponential fit to the $p_{T}$ spectra.} \label{table}
\vspace{+0.5cm}
\end{table}

\section{Conclusion}

The $\Lambda$(1520)/$\Lambda$ ratio in d+Au collisions at
$\sqrt{s_{\rm NN}} = $ 200 GeV is in agreement with the ratio in
p+p collisions. This indicates a zero life time between the
chemical and thermal freeze-out where elastic interactions occur.
The slopes of the momentum spectrum of the $\Lambda$(1520)
measured in p+p and d+Au is equal, which also leads to the
conclusion of a no hadronic phase after chemical freeze-out.



\end{document}